\documentclass[11pt]{article}
\usepackage{amsmath,amssymb}

\def\g{\mathbf{g}}
\def\t{\mathbf{t}}
\def\diag{\mathrm{diag}\,}
\def\id{\mathbf{1}}
\def\J{\mathcal{J}}
\def\P{\mathcal{P}}
\def\Q{\mathcal{Q}}
\def\N{\mathcal{N}}
\def\D{\mathcal{D}}
\def\ap{\vec a_{\P_1}}
\def\bp{\vec b_{\P_1}}
\def\bq{\vec b_{\Q_1}}
\def\aq{\vec a_{\Q_1}}
\def\sech{\mathrm{sech}\,}

\begin{document}
\begin{titlepage}
\begin{flushright}
{\bf \today} \\
                 DAMTP-2010-13\\
                 \end{flushright}
                 \begin{centering}
                 \vspace{.2in}

{\large {\bf The Force Between Giant Magnons }}

\vspace{.3in}

                 Nick Dorey and Rui F. Lima Matos\\
                 \vspace{.2 in}
                 DAMTP, Centre for Mathematical Sciences \\
                         University of Cambridge, Wilberforce Road \\
                         Cambridge CB3 0WA, UK \\

                         \vspace{.4in}

{\bf Abstract} \\

\end{centering}
We compute the force and torque between well-separated, slowly-moving 
Giant Magnons with arbitrary orientations on $S^{5}$. We propose an
effective Hamiltonian for Giant Magnons in this regime.

\end{titlepage}

\section{Introduction}

An interesting feature of field theory is that classical soliton
solutions give rise to particle-like states after quantisation
\cite{GJ, DHN, FK}.  The resulting particles can often be usefully
described by an effective Lagrangian or Hamiltonian.  Famous examples
of this phenomenon occur in the context of sine-Gordon Thirring
duality \cite{Cn}, Olive-Montonen duality of ${\cal N}=4$ SUSY
Yang-Mills theory \cite{MontO} and the role of Skyrmions as baryons in
large-$N_{c}$ QCD \cite{ANW}. An interesting recent example has
emerged in the context of the AdS/CFT correspondence where certain
solitons on the worldsheet theory of the string on $AdS_{5}\times
S^{5}$, known as Giant Magnons \cite{HM}, are dual to the elementary
excitations of the gauge theory spin chain. Finding an effective
Hamiltonian description of these solitons is an interesting
problem which should be important in the context of string
quantisation. In the case of uncharged Giant Magnons moving 
on the same $S^{2}$ submanifold of $S^{5}$, the dynamics is closely 
related to that of sine-Gordon solitons 
and an effective Hamiltonian was proposed in \cite{AJ}.  
Here we will address this problem in the more general case of charged   
Giant Magnons with arbitrary orientation on $S^{5}$. However we will
solve the problem only in the case where the Giant Magnons are 
well-separated and slowly moving.  

Here we will employ a simple approach to the problem due to Manton
\cite{Manton} which is available only for integrable field equations. 
We illustrate the method using the sine-Gordon equation, 
\begin{eqnarray}
 \frac{\partial^{2} \phi}{\partial t^{2}} - 
\frac{\partial^{2} \phi}{\partial x^{2}} & = & -\frac{m}{\beta} \sin
\beta\phi \nonumber \end{eqnarray}
as an example. This has a soliton solution of the form, 
\begin{eqnarray}
\phi_{K}(x,t) & = & 4\tan^{-1}\left[\exp(\gamma(v)m(x-vt))\right]
\nonumber \end{eqnarray}
where $\gamma(v)=1/\sqrt{1-v^{2}}$ is the Lorentz factor. The soliton
or kink interpolates between the vacuum values $\phi=0$ and
$\phi=2\pi/\beta$ and has a well defined position $X(t)=vt$ where the
field takes the midpoint value $\phi=\pi/\beta$. The kink moves at
constant velocity $v$ and behaves like a relativistic 
particle of mass $M=8m/\beta$. A corresponding
anti-kink solution is obtained by $\phi \rightarrow -\phi$ with an
appropriate choice for the branch of the inverse tangent.

We now consider a configuration which contains both a kink and an
anti-kink. A static configuration of this sort only solves the
equation of motion in the limit of infinite separation. However, the
sine-Gordon equation, being integrable, 
has an exact solution $\phi_{K\bar{K}}$ describing the scattering
of a kink with an anti-kink. In the center of mass frame this is given
as; 
\begin{eqnarray}
\tan \left(\frac{\beta\phi_{K\bar{K}}}{4}\right) & =  & 
\frac{1}{v}\,\frac{\sinh(\gamma(v)vmt)}{\cosh(\gamma(v)vx)}  \nonumber 
\end{eqnarray}
This solution asymptotes to a linear superposition of a kink and
anti-kink, with velocities $\pm v$ respectively, at very early and
late times 
$t\rightarrow \pm \infty$. As before we define the worldline of the
kink as a 
solution $x=X(t)$ of $\phi(x,t) =
\pi/\beta$ or,  
\begin{eqnarray} v \cosh(\gamma(v)vX) & =& \sinh(\gamma(v)vmt) 
\label{wl}
\end{eqnarray} 
For non-relativistic velocities $v<<1$, (\ref{wl}) has an approximate
solution at late times of the form,
\begin{eqnarray}
X(t) & = & -\frac{1}{m}\log v
\,\,+\,\,vt\,\,-\,\,\frac{1}{m}\exp(-2vmt) \nonumber 
\end{eqnarray}
The first two terms correspond to the asymptotic free motion of the
kink. The final term gives rise to an exponentially small 
acceleration, 
\begin{eqnarray}
\ddot{X} & = & -4m\,\exp(-2m X)  
\label{attr}
\end{eqnarray}
which can be interpreted in terms of an attractive potential between
the kink and anti-kink, separated by a distance $s=2X$, 
\begin{eqnarray}
\mathcal{V}(s) & = & -\frac{32m}{\beta^{2}}\, \exp(-ms) 
\label{vsg}
\end{eqnarray}
or force given by $F=-\partial\mathcal{V}/\partial s$. 
Finally the effective Hamiltonian describing the slow motion of both
the kink and an anti-kink with positions $X_{K}$ and $X_{\bar{K}}$
respectively and
conjugate momenta $P_{i}=M\dot{X}_{i}$ for $i=K,\bar{K}$ takes the form, 
\begin{eqnarray}
\mathcal{H}_{K\bar{K}} & = & \frac{P_{K}^{2}}{2M}\,\,+\,\,  
\frac{P_{\bar{K}}^{2}}{2M}\,\,+\,\,\mathcal{V}\left(|X_{K}-X_{\bar{K}}|\right)
\nonumber 
\end{eqnarray}
with potential $\mathcal{V}$ given in (\ref{vsg}). This result agrees
with other approaches to determining the effective Hamiltonian at
large separation. The method of \cite{BS,BS2} involves studying the motion of
poles of the analytically continued energy density\footnote{This
  approach does not yield a Hamiltonian description of the motion for
  generic separations \cite{RS}.}. One may also
proceed by constructing static multi-soliton solutions by allowing 
discontinuities on the spatial derivatives of the
fields across the particle positions 
\cite{PS,Rajaraman}. While the first of these requires the
knowledge of explicit mulit-soliton solutions, the second method does
not, thus being also applicable to non-integrable theories.
Finally, for this system there is an exact Hamiltonian description 
given by the two-body Ruijsenaars-Schneider model \cite{RS}, 
with a Hamiltonian 
which reduces to (\ref{vsg}) for low velocity and large 
separation\footnote{Note however that 
the naive definition (\ref{wl}) of the soliton worldline used
above will not reproduce the correct Hamiltonian description except in this
regime.} $|X_{K}-X_{\bar{K}}|>> m^{-1}$.    

In this paper we will perform an analogous calculation to derive an
effective Hamiltonian for two well-separated, slowly moving Giant Magnons.  
In addition to linear momenta $P^{(1)}$ and $P^{(2)}$, canonically
conjugate to the soliton positions $X^{(1)}$ and $X^{(2)}$, 
these solitons also 
carry conserved $U(1)$ charges $Q^{(1)}$ and $Q^{(2)}$ corresponding to
internal motion with respect to angular coordinates $\Theta^{(1)}$ and
$\Theta^{(2)}$. Each soliton also has additional coordinates
describing its orientation on $S^{5}$. Starting from an appropriate 
two-soliton solution of the string $\sigma$-model,  we 
compute the force and torque for dyonic giant magnons, each with
arbitrary polarisations in $S^5$, in the large-separation regime where 
$|\Delta X|\equiv|X^{(2)}-X^{(1)}|>>1$, and where both linear and internal
soliton motion is slow. These latter conditions correspond to momenta 
$P^{(1)}$, $P^{(2)}$ and  charges $Q^{(1)}$, $Q^{(2)}$ of order one
in the regime of large 't Hooft coupling $\lambda=g^{2}_{YM}N>>1$. 
The resulting effective Hamiltonian takes the form:
\begin{equation}
\mathcal{H}=2M-\frac{[P^{(1)}]^2}{2M}-\frac{[P^{(2)}]^2}{2M}+
\frac{[Q^{(1)}]^2}{2M}+
\frac{[Q^{(2)}]^2}{2M}+\mathcal{V}(|\Delta
X|,\Theta^{(1)},\Theta^{(2)}),
\label{res1}
\end{equation}
where $M=\sqrt{\lambda}/\pi$ is the mass of each soliton
and the two-body potential is shown to have the form, 
\begin{equation}\label{eq:VV}
\mathcal{V}(|\Delta X|,\Theta^{(1)},\Theta^{(2)})=\frac{\sqrt{\lambda}}{\pi}
f(\Theta^{(1)},\Theta^{(2)})\, e^{-|\Delta X|},
\end{equation}
\begin{equation}
f(\Theta^{(1)},\Theta^{(2)})=K^+
\cos(\Theta_1+\Theta_2)+K^-\cos(\Theta_1-\Theta_2),\
\end{equation}
\begin{equation}
K^\pm=K^\pm\biggl(\|\vec J_1\|,\|\vec J_2\|,\vec J_1\cdot\vec
J_2\biggr).
\end{equation}
The coefficients $K^\pm$ are given explicitly in (\ref{eq:Ks}) 
as functions of the $SO(4)$ invariants formed from 
the $S^{5}$ angular momenta $\vec J_1$ and $\vec J_2$ of
the two Giant Magnons. 

For solitons of charge $Q^{(1)}$, $Q^{(2)}=0$ the two-body potential
(\ref{eq:VV})
reduces to the sine-Gordon result (\ref{vsg}). This reflects the
correspondence between uncharged Giant Magnons and sine-Gordon kinks 
induced by Pohlmeyer reduction \cite{Pohl} and our results are
consistent with those of \cite{AJ} in this case. The effective Hamiltonian for 
charged Giant Magnons with parallel angular momenta on $S^{5}$ was
also considered in \cite{DHM}. In particular, general constraints on
the Hamiltonian were considered and a toy model which
correctly reproduces the known analytic structure of the magnon
S-matrix was proposed in this reference. Restricting our result to the
case of parallel $S^{5}$ angular momenta 
we find that the Hamiltonian (\ref{res1}) 
has the same asymptotic form as the toy model considered
in \cite{DHM}. We believe our result should provide a useful starting
point for the search for an exact Hamiltonian description of Giant
Magnons.

\section{Giant Magnons on $\mathbb{R}\times S^5$}

The bosonic sector of type IIB superstring theory in the
background $AdS_5\times S^5$ can be described by the following
$\sigma$-model action
\begin{equation}\label{eq:action}
S=\frac{\sqrt{\lambda}}{2\pi}\int dx\,dt\,\biggl\{
\overbrace{\eta^{ab}\partial_a Y^\mu\partial_b Y_\mu+
\Lambda_1(Y_\mu Y^\mu+1)}^{AdS_5}
+
\underbrace{\eta^{ab}\partial_a X^j\partial_b X_j+
\Lambda_2(X_j X^j-1)}_{S^5}\biggr\},
\end{equation}
where $(x,t)$ are the string worldsheet coordinates,
$\vec Y,\vec X\in\mathbb{R}^6$ parametrise the string embedding,
\begin{align}\label{eq:parameterisationAdS}
Y_\mu Y^\mu &= -Y_{-1}^2-Y_0^2+Y_1^2+Y_2^2+Y_3^2+Y_4^2=1,\\
X_j X^j &= X_1^2+X_2^2+X_3^2+X_4^2+X_5^2+X_6^2=1,
\label{eq:paremeterisationS}
\end{align}
$\eta^{ab}$ is the (Minkowskian) string worldsheet metric in conformal
gauge and $\Lambda_{1,2}$ are Lagrange multipliers that enforce
(\ref{eq:parameterisationAdS}) and (\ref{eq:paremeterisationS}).
In what follows it will be convenient to introduce the following
parametrisation of
$S^5$,
\begin{equation}\label{eq:Zs}
Z_1=X_1+iX_2,\quad Z_2=X_3+iX_4,\quad
Z_3=X_5+iX_6,\quad\text{with}\quad
\sum_{j=1}^3|Z_j|^2=1.
\end{equation}

The action (\ref{eq:action}) has the global symmetry $SO(4,2)\times SO(6)$.
Through
this paper we will be interested in a subset of the solutions of the
Euler-Lagrange equations for the above action, namely those
restricted to the submanifold 
$\mathbb{R}\times S^5\subset AdS_5\times S^5$. We will
choose the gauge $Y^0=t$. The generator of time translations will
then be given by
\begin{equation}
\label{eq:delta}
\Delta=i\frac{\sqrt\lambda}{2\pi}\int dx\,\dot
Y_0=i\frac{\sqrt\lambda}{2\pi}\int dx,
\end{equation}
and the generator of rotations in $S^5$ by
\begin{equation}
\label{eq:Jij}
J_{ij}=i\frac{\sqrt\lambda}{2\pi}\int dx\,X_{[i}\dot X_{j]},\quad
i,j=1,\dots,6.
\end{equation}

Following \cite{HM}, 
we will consider the limit when the angular momentum $J\equiv J_{12}$
is very large and look for solutions for which the difference
$E\equiv\Delta-J$ remains finite \cite{HM}. In this limit the
string becomes infinitely long and we can relax the closed string
boundary condition, allowing different values for the fields $Z_j$ at
left and right infinity $x\rightarrow\pm\infty$.
The simplest solution one can construct is the BMN groundstate
solution \cite{BMN}  
\begin{equation}
\label{eq:BMN}
Z^{(BMN)}_{1}=e^{it},\quad Z^{(BMN)}_2=Z^{(BMN)}_3=0.
\end{equation}
which describes the motion of a point-like string
travelling along the equator $S^1$ of $S^5$ with very large angular
momentum $J$. It saturates a BPS bound obeying, 
\begin{equation}
\label{eq:E_BMN}
\Delta-J=0.
\end{equation}

The Giant Magnon  
is a solitonic excitation of the BMN 
vacuum (\ref{eq:BMN}). The simplest case is the original uncharged
Giant Magnon of \cite{HM} which moves on and $S^{2}\in S^{5}$. The
solution takes the form,  
\begin{equation}
\begin{cases}
Z_1=e^{it}\biggl\{\gamma^{-1}\tanh\biggl(\gamma[(x-x_0)-V\,t]
\biggr)\biggr]-iV\biggr\},\\
Z_2=\gamma^{-1}\,\sech\biggl(\gamma[(x-x_0)-V\,t]\biggr),\\
Z_3=0,
\end{cases}
\label{c1}
\end{equation}
with $V=\cos(p/2)$ and $\gamma=(1-V^2)^{-\frac12}$.
The solution carries a conserved momentum $p$ which is related to the
large-$x$ asymptotics of the fields according to,  
\begin{equation}
\label{eq:lim_OGM}
\displaystyle\lim_{x\to\pm\infty}Z_j[x,t;p,x_0]=-i\,Z_j^{(BMN)}[t]\,e^{\pm
i\frac{p}2},\quad j=1,2,3.
\end{equation}
This excitation contributes a finite amount to the anomalous dimension
$\Delta-J$ and has dispersion, 
relation, 
\begin{equation}
\label{eq:E_OGM}
\Delta-J=\frac{\sqrt\lambda}{\pi}\sin\biggl(\frac{p}2\biggr),
\end{equation}
The solution (\ref{c1}) has a well-defined center, 
\begin{equation}
X(t)=x_0+V\,t.
\end{equation}
which moves with constant velocity $V$. The center $X(t)$ is the point where 
$|Z_2|$ is maximal which therefore solves, 
\begin{equation}\label{eq:X}
\partial_x |Z_2[X(t),t;p,x_0]|=0.
\end{equation}
We will adopt this definition for the position coordinate of the
soliton in a more general context below. The conserved
momentum $p$ introduced above is canonically conjugate to $X$
\cite{HM} and 
the pair $\{p,X\}$ can be regarded as the collective coordinates of this
neutral Giant Magnon solution.

The above solution admits a generalisation which, in addition to
linear momentum, carries a conserved $U(1)$ charge.  This solution is
the dyonic giant magnon of \cite{CDO} (see also \cite{SV})) which corresponds to a string
moving on an $S^{3}$ submanifold of $S^{5}$,
\begin{equation}
\label{eq:DGM_sol_simple}
\begin{cases}
Z_1=e^{it}\biggl\{\sin\bigl(\frac{p}2\bigr)
\tanh\,u(x-x_0,t;r,p)-i\cos\bigl(\frac{p}2\bigr)\biggr\},
\\
Z_2=\sin\bigl(\frac{p}{2}\bigr)
\,\mathrm{sech}\,u(x-x_0,t;r,p)\,e^{i[v(x-x_0;r,p)+\theta_0]},
\\
Z_3=0.
\end{cases}
\end{equation}
Here we have introduced
the functions
\begin{align}\label{eq:u}
u(x,t;r,p)&=\frac{2r(1+r^2)\sin\bigl(\frac{p}2\bigr)}
{1-2r^2\cos\,p+r^4}[x-V\,t],
\\
\label{eq:v}
v(x,t;r,p)&=\frac{2r(r^2-1)\cos\bigl(\frac{p}2\bigr)}
{1-2r^2\cos\,p+r^4}[x-V^{-1}\,t],
\\
\label{eq:V}
V&=\frac{2r}{r^2+1}\cos\bigl(\frac{p}2\bigr).
\end{align}
The conserved $U(1)$ charge is, 
$Q=\frac{\sqrt\lambda}{2\pi}(r-r^{-1})$, $(r\in\mathbb R)$,
Setting $Q=0$ we return to the neutral Giant Magnon described in the
preceding paragraph. The dispersion relation of the new solution is given by
\begin{equation}
\label{eq:E_DGM}
\Delta-J=\sqrt{Q^2+\frac{\lambda}{\pi^2}\sin\biggl(\frac{p}2\biggr)}.
\end{equation}
Its asymptotics are identical to (\ref{eq:lim_OGM}),
\begin{equation}
\label{eq:lim_DGM}
\displaystyle\lim_{x\to\pm\infty}Z_j[x,t;p,x_0,Q,\theta_0]=
-i\,Z_j^{(BMN)}[t]\,e^{\pm
i\frac{p}2},\quad j=1,2,3.
\end{equation}

In addition to a position $X(t)$ defined as above, the 
dyonic Giant Magnon has an additional degree of freedom $\Theta(t)$ 
corresponding to $U(1)$ rotations acting on the target space
coordinate $Z_{2}$. Here $\Theta(t)$ is an angular coordinate which 
evolves linearly on the one-soliton solution, 
\begin{equation}
\Theta(t)=\theta_0+\omega\,t,
\end{equation}
with angular velocity, 
\begin{equation}
\omega = \frac{1-r^2}{1+r^2}.
\end{equation}
and is canonically conjugate to the $U(1)$ charge $Q$
The initial value $\theta_0$ corresponds to the argument of
$Z_2$ at time $t=0$ evaluated at the point $x=x_0$. More generally 
we can define that phase at any other time $t$ by
\begin{equation}\label{eq:Theta}
\Theta(t)=\arg\biggl\{Z_2[X(t),t;p,x_0,Q,\theta_0]\biggr\}.
\end{equation}
The set of collective coordinates for a dyonic Giant magnon are therefore 
$\{p,X;Q,\Theta\}$.

The dyonic Giant Magnon solution involves the choice
of an $S^3\subset S^5$. For the solution described above this
corresponds to the choice $Z_3=0$. 
We obtain more general solutions which asymptote to the same vacuum 
at spatial infinity by performing an $SO(4)$ rotation on
the embedding coordinates $(X_3,X_4,X_5,X_6)$. Of the possible
$SO(4)$ rotations, a rotation acting only on $(X_5,X_6)$ - 
i.e. a phase shift in $Z_3$ - will
not change our initial
solution. Similarly a rotation in the
plane $(X_3,X_4)$ will simply induce a phase shift in $Z_2$, that can
be absorbed by shifting the constant $\theta_0$. Thus the stabilizer
in $SO(4)$ of the solution (\ref{eq:DGM_sol_simple}) is $SO(2)\times SO(2)$. 
Inequivalent embeddings of the dyonic Giant Magnon on $S^{5}$ are
therefore parametrised by the coset \cite{SV, CSV},
\begin{equation}
Gr_2(\mathbb R^4)=\frac{SO(4)}{SO(2)\times SO(2)}\approx \frac{SO(3)\times
SO(3)}{SO(2)\times SO(2)}\approx S_+^2\times S_-^2.
\end{equation}
We introduce polar coordinates on each of the spheres
$S_\pm^2$, by choosing a unit vectors,
\begin{equation}
\vec n_\pm = (\cos[\psi_+\mp\psi_-]
\sin\alpha_\pm,\sin[\psi_+\mp\psi_-]
\sin\alpha_\pm,\cos\alpha_\pm).
\end{equation}
Finally, performing an $SO(4)$ rotation on (\ref{eq:DGM_sol_simple}), 
the most general Giant Magnon solution 
solution in $\mathbb R\times S^5$ has then the
form \cite{CSV},
\[Z_j=Z_j[x,t;p,x_0,Q,\theta_0,\vec n_\pm],\quad j=1,2,3,\]
with,
\begin{equation}
\label{eq:DGM_sol}
\begin{cases}
Z_1=e^{it}\biggl\{\sin\bigl(\frac{p}2\bigr)
\tanh\,u(x-x_0,t;r,p)-i\cos\bigl(\frac{p}2\bigr)\biggr\},
\\
\begin{split}
Z_2=\,e^{i\psi^+}\sin\bigl(\frac{p}{2}\bigr)\mathrm{sech}\,u(x-x_0,t;r,p)\,[
\cos\{v(x-x_0;r,p)+\theta_0\}\cos\Delta\alpha
\\
+i\sin\{v(x-x_0;r,p)+\theta_0\}\cos\bar\alpha],
\end{split}
\\
\begin{split}
Z_3=\,e^{i\psi^-}\sin\bigl(\frac{p}{2}\bigr)\mathrm{sech}\,u(x-x_0,t;r,p)\,[
\cos\{v(x-x_0;r,p)+\theta_0\}\sin\Delta\alpha
\\
+i\sin\{v(x-x_0;r,p)+\theta_0\}\sin\bar\alpha],
\end{split}
\end{cases}
\end{equation}
where $\bar\alpha=\alpha_+-\alpha_-$,
$\Delta\alpha=\alpha_++\alpha_-$.
This solution has the asymptotics
(\ref{eq:lim_DGM}) and it obeys the dispersion relation (\ref{eq:E_DGM}), with
$Q=\mp\mathrm{tr}[(\vec J^\pm)^2]$. The angular momentum of the
solution can be written in the form
\begin{align}\label{eq:ang_mom}
         \vec J &\equiv (J_{ij})=
\sum_{{a\in\{1,2,3\}}\atop{k=\pm}}\,J_a^k\t^k_a,\quad i,j=3,\dots,6\\
         \vec J^\pm &\equiv (J_a^\pm)=\pm
Q(\cos(2\alpha_\pm),-\cos[\psi_+\mp\psi_-]\sin(2\alpha_\pm),\sin[\psi_+\mp\psi_-]\sin(2\alpha_\pm)),
\end{align}
with $\{\t_a^\pm\}$ a basis
for the Lie algebra
$\mathfrak{so}(4)=\mathfrak{su}(2)\oplus\mathfrak{su}(2)$ which is
given in the Appendix.
As expected, $\vec J^+$ and $\vec J^-$ parametrize two
copies of $S^2$ with the same fixed radius equal to the $U(1)$ charge
$Q$. The complete set of collective coordinates of the most general
Giant Magnon solution is therefore 
$\{p,X,Q,\Theta,\vec n_\pm\}$. As above, $X(t)$ and $\Theta(t)$ evolve 
linearly according to, 
\begin{equation}
X(t)=x_0+V[p]\,t,\quad \Theta(t)=\theta_0+\omega[Q]\,t
\end{equation}
\begin{equation}\label{eq:omega}
V[p]=\frac{\sqrt{\lambda}}{\pi}
M(r)^{-1}\cos\biggl(\frac{p}2\biggr),\quad\omega[Q]=\frac{Q}{M(r)},\quad
M(r)\equiv
E[p=\pi]=\frac{\sqrt{\lambda}}{2\pi}\frac{1+r^2}{r},
\end{equation}
The remaining collective coordinates are constants of motion. 
The quantity $M(r)$ 
will play the role of the soliton mass in the non-relativistic limit. 


\section{The Force and Torque between Giant Magnons}
Let us now consider a configuration containing two Giant Magnons
 of the form (\ref{eq:DGM_sol}),
\begin{equation}\label{eq:DGMs}
\begin{split}
Z_{j}^{(1)}\equiv
Z_j[x,t;p^{(1)},x_{0}^{(1)},Q^{(1)},\theta_{0}^{(1)},\vec n_{\pm}^{(1)}],
\\
Z_{j}^{(2)}\equiv
Z_j[x,t;p^{(2)},x_{0}^{(2)},Q^{(2)},\theta_{0}^{(2)},\vec n_{\pm}^{(2)}],
\end{split}
\end{equation}
separated by
a distance $d=|x_{0}^{(1)}-x_{0}^{(2)}|$, at time $t=0$, and with
momenta $p^{(1)}=\pi-\delta$ and
$p^{(2)}=\pi+\delta$ in the centre of momentum frame and
charges
$Q^{(1)}=\sqrt{\lambda}(r_1-r^{-1}_1)/2\pi$ and
$Q^{(2)}=\sqrt{\lambda}(r_2-r^{-1}_2)/2\pi$. 
The effective Hamiltonian will be invariant under all global
symmetries left unbroken by the BMN vacuum.  
Thus, without loss of generality we may fix 
$x_{0}^{(1)}=0$ and $x_{0}^{(2)}=-d$
and, by using the global $SO(4)$ symmetry, fix $\vec n_{\pm}^{(1)}$ of the
first
soliton to be $(0,0,1)$. In other words we can choose, 
\begin{equation}
\label{eq:DGMs_par}
\begin{cases}
\theta_{0}^{(1)}=0,\quad\alpha_{\pm}^{(1)}=0,\quad\psi_{\pm}^{(1)}=0,\\
\theta_{0}^{(2)}=\theta_{0},\quad\alpha_{\pm}^{(2)}=\alpha_{\pm}\quad\psi_{\pm}^{(2)}=\psi_{\pm}.
\end{cases}
\end{equation}
Using the stabiliser of the first soliton solution we may further
rotate the second soliton in the $Z_3$ plane and set
$\psi_-=0$. 

As for the sine-Gordon case discussed in the introduction, a linear
superposition of two solitons does not solve the equations of motion
except in the limit of infinite separation. 
To compute the force between these solitons we
need to find an exact scattering solution  $\mathcal Z_j$ which asymptotes to a
configuration of two well-separated solitons, with parameters chosen as 
in (\ref{eq:DGMs_par}) at very early and late times. By shifting the
time coordinate we can choose the origin $t=0$ to lie in the early
asymptotic region and demand, 
\begin{equation}\label{eq:sum}
         \mathcal Z_j\xrightarrow{t=0,d\rightarrow\infty}
         Z_j^{(1)}+Z_j^{(2)}.
\end{equation}


The equations of motion of the $AdS_{5}\times S^{5}$ $\sigma$-model
are integrable and it is possible to construct the required two soliton 
scattering solution explicitly 
using the dressing method \cite{SV, CSV}. This is accomplished in the
Appendix where the resulting solution is specified by equation 
(\ref{eq:polarizations_corrected}).
As expected, near $x=0$ and $x=-d$,
$\mathcal Z_j$ resembles $Z_j^{(1)}$ and $Z_j^{(2)}$
respectively. Expanding the solution around this limit 
to first non-trivial order we obtain, 
\begin{equation}\label{eq:Z_corrections}
\begin{cases}
\mathcal{Z}_j\approx Z_j^{(1)}+\delta
Z^{(1)}_j[x,t;\delta,Q^{(1)},Q^{(2)},\theta_0,\alpha_\pm,\psi_+]\,
e^{-D_2},\quad\text{for }
|x-\omega^{(1)}t|\sim e^{-D_2},\\
\mathcal{Z}_j\approx
Z_j^{(2)}+\delta
Z^{(2)}_j[x,t;\delta,Q^{(1)},Q^{(2)},\theta_0,\alpha_\pm,\psi_+]\,
e^{-D_1},\quad\text{for }
| x+d-\omega^{(2)}t|\sim e^{-D_1},
\end{cases}
\end{equation}
where 
\begin{equation}
D_j\equiv
u(d,0;r_j,p^{(j)})=\frac{2r_j(1+r_j^2)}{1-2r_j^2\cos\,p^{(j)}+r_j^4}\,d,\quad
j=1,2.
\end{equation}

The correction terms $\delta Z^{(1)}_{j}$ and 
$\delta Z^{(2)}_{j}$, will
alter the linear evolution of the collective coordinates for each
soliton. We make the ansatz, 
\begin{equation}\label{eq:XXThetaTheta}
\begin{cases}
X^{(1)}(t)=V^{(1)}\,t+\delta X^{(1)}(t)\,e^{-D_2},\quad
\Theta^{(1)}(t)=\omega^{(1)}\,t+\delta\Theta^{(1)}(t)\,e^{-D_2}\\
X^{(2)}(t)=x_0+V^{(2)}\,t+\delta X^{(2)}(t)\,e^{-D_1},\quad
\Theta^{(2)}(t)=\theta_0+\omega^{(2)}\,t+\delta\Theta^{(2)}(t)\,e^{-D_1}.
\end{cases}
\end{equation}
For computing the two-body force and torque, 
we need to compute the corrections $\delta X^{(1)}(t)$ and 
$\delta\Theta^{(1)}(t)$ to the world-line of the first soliton.
At early times $t\simeq 0$, the 
position coordinate of the soliton is determined as in
(\ref{eq:X}) by, 
\begin{equation}
0=\partial_x|\mathcal Z_2|\approx
\partial_x\biggl\{|Z_2^{(1)}|\biggl(1+\Re\biggl[\frac{\delta
Z_2^{(1)}}{Z_2^{(1)}}\biggr]\,e^{-D_{2}}\biggr)\biggr\},\quad x\approx
\omega^{(1)}\,t+\delta X^{(1)}\,e^{-D_{2}},
\end{equation}
As in the sine-Gordon case studied in the introduction we expect this
simple-minded definition of the worldline to be adequate only in the
limit of large separation: $d\rightarrow\infty$.

The correction $\delta
Z^{(1)}_2$  can expressed in terms of the functions
\begin{equation}
u_a\equiv u(x,t,r_j,p^{(a)}),\quad
v_a\equiv v(x,t,r_1,p^{(a)}),\quad a=1,2,
\end{equation}
defined for each soliton as in Equation (\ref{eq:v}, \ref{eq:V}) above.
The lengthy explicit expression for $\delta Z^{(1)}_{2}$
is given in Equation (\ref{eq:deltaZ2}) of the Appendix.
Taking the non-relativistic limit of low velocity $\delta<<1$, 
we then find,
\[\partial_x|\mathcal Z_2|
=(u_1'\partial_{u_1}+u_2'
\partial_{u_2}+v_1'\partial_{v_1}+v_2'\partial_{v_2})|\mathcal
Z_2|\approx
(u_1'\partial_{u_1}+u_2'\partial_{u_2})|\mathcal Z_2|,
\]
where prime denotes partial derivative with respect to the spacelike
worldsheet coordinate $x$ and we have neglected the contribution
comming from $v'_j$ because these are of order $\delta$. From the
explicit expression (\ref{eq:deltaZ2}) for $\delta Z_2^{(1)}$  and
(\ref{eq:frac}) we have
that, 
\[\partial_{u_2}\Re\biggl[\frac{\delta
Z_2^{(1)}}{Z_2^{(1)}}\biggr]_{x=\omega^{(1)}t}=-\Re\biggl[\frac{\delta
Z_2^{(1)}}{Z_2^{(1)}}\biggr]_{x=\omega^{(1)}t}=0,\]
thus deriving, in the non-relativistic limit,
\begin{equation}\label{eq:X_eq}
\delta X^{(1)}\approx\frac{1}{u_1'}\Re\biggl[\partial_{u_1}\biggl(\frac{\delta
Z^{(1)}_2}{Z_2^{(1)}}\biggr)\biggr]_{x=\omega^{(1)}t}.
\end{equation}

As in equation (\ref{eq:Theta}) above, we define the internal angle 
$\Theta^{(1)}(t)$ of the first soliton as the phase of $Z_{2}$
evaluated at $x=X^{(1)}(t)$.  
In a similar fashion, by expanding the argument of $\mathcal Z_2$ in
the neighbourhood of $x=\omega^{(1)} t$, for small fixed $t$, we get
\begin{equation}
\label{eq:Theta_eq}
\delta\Theta_1^{(1)}(t)=\Im\biggl\{\frac{\delta
Z_2^{(1)}}{Z_{2}^{(1)}}\biggr\}_{x=\omega^{(1)}\,t}.
\end{equation}
We then find, using (\ref{eq:X_eq}) and (\ref{eq:Theta_eq}) with $\delta Z^{(1)}_2$ given by
(\ref{eq:deltaZ2}) and by taking the non-relativistic limit
$\delta<<1$, that
\begin{multline}
\label{eq:Norm_sol}
\delta X^{(1)}(t)=- \frac{r_1^2+1}{r_1}e^{-(u_2+D)}
\biggl\{\frac{r_1^2+r_2^2}{(r_1+r_2)^2}\cos\alpha_+\cos\alpha_-
\cos(v_1-v_2-\theta_0-\psi_+)
\\
+\frac{1+r_1^2r_2^2}{(1+r_1
r_2)^2}\sin\alpha_+\sin\alpha_-\cos(v_1+v_2+\theta_0-\psi_+)\biggr\},
\end{multline}
\begin{multline}
\label{eq:Theta_sol}
\delta\Theta_1^{(1)}(t)=4\,r_1r_2\, e^{-(u_2+D)}
\biggl\{\frac{1}{(r_1+r_2)^2}\cos\alpha_+\cos\alpha_-
\cos(v_1-v_2-\theta_0-\psi_+)
\\
+
\frac{1}{(1+r_1
r_2)^2}\sin\alpha_+\sin\alpha_-\cos(v_1+v_2+\theta_0-\psi_+)\biggr\},
\end{multline}
If we further impose that the two solitons have charges
$Q^{(1)},Q^{(2)}$ of order one for $\lambda>>1$, 
or equivalently that the angular velocities are
small, $\omega^{(j)}<<1$, we find
that
\begin{align}
m^{(1)}\ddot
X^{(1)}
&\approx-\frac{\sqrt{\lambda}}\pi\biggl\{K_+\cos(\Theta^{(1)}+\Theta^{(2)}-\psi_+)+K_-\cos(\Theta^{(1)}-\Theta^{(2)}-\psi_+)\biggr\}\,e^{(X^{(2)}-X^{(1)})},
\\
m^{(1)}\ddot
\Theta^{(1)}
&\approx\frac{\sqrt{\lambda}}\pi\biggl\{K_+\cos(\Theta^{(1)}+\Theta^{(2)}-\psi_+)+K_-\cos(\Theta^{(1)}-\Theta^{(2)}-\psi_+)\biggr\}\,e^{(X^{(2)}-X^{(1)})},
\end{align}
with the factors $K_\pm=K_\pm(\vec J_1^\pm,\vec J_2^\pm)$ depending only on
the angular momenta $\vec J_1,\vec J_2$ of the two solitons:
\begin{equation}\label{eq:Ks}
K_\pm = -\frac{2}{J_1J_2}\frac{(J_1\pm J_2)^2}
{\|\vec J_1^--\vec J_2^-\|
\|\vec J_1^+-\vec J_2^+\|}
\sqrt{(J_1J_2\mp \vec J_1^-\cdot \vec J_2^-)(J_1J_2\mp \vec J_1^+\cdot
\vec J_2^+)}
\end{equation}

By defining the effective potential
\begin{equation}
\mathcal V(x,\theta_1,\theta_2) =
\frac{\sqrt{\lambda}}\pi\biggl\{K_+\cos(\theta_1+\theta_2)+K_-\cos(\theta_1-\theta_2)\biggr\}\,e^{-x},
\end{equation}
we can write for the force $\mathcal F^{(1)}$ and torque $\mathcal
T^{(1)}$  exerted by the second soliton on the first,
\begin{align}\label{eq:Force}
\mathcal F^{(1)}\equiv m_1 \ddot X^{(1)} &\approx \frac{d}{d
X^{(1)}}\mathcal V(|\Delta X|,\Theta_1,\Theta_2),\\
\label{eq:Torque}
\mathcal T^{(1)}\equiv m_1 \ddot \Theta_1 &\approx -\frac{d}{d
\Theta_1}\mathcal V(|\Delta X|,\Theta_1,\Theta_2),
\end{align}
with $\Delta X=X^{(2)}-X^{(1)}$ and where we have absorbed $\psi_+$ in the
initial value of $\Theta^{(1)}$.
Identical results are obtained for the force and torque on the
second soliton. 

For neutral Giant Magnons ($Q^{(1)}=Q^{(2)}=0$) the force reduces
to
\begin{equation}
\mathcal{F}^{(1)}\approx
-4\frac{\sqrt{\lambda}}{\pi}\cos(\Delta\alpha)\cos(\psi_+)\,e^{-|\Delta
X|},
\end{equation}
where $\Delta\alpha = \alpha_+-\alpha_-$. Here $\psi_+$ is the phase 
difference between two solitons embedded on the same
$\mathbb R\times S^2$ when we make $\alpha_\pm=0$ ($\psi_+=0$
for a pair of solitons and $\psi_+=\pi$ for a soliton-antisoliton
pair). If both
solitons have the same phase, the equation of motion is simply
\begin{equation} \label{eq:ForceGM}
         \ddot X^{(1)} = -4 e^{-|\Delta X|}.
\end{equation}
We can compare this result with the sine-Gordon result (\ref{attr})
which can be written as, 
\begin{equation}\label{eq:sg}
         \ddot X^{(1)}_{sG} = -4m\,e^{-m|\Delta X|},
\end{equation}  
As expected the two results agree when we set the sine-Gordon mass $m$
to unity. 


By expanding their exact dispersion relation (\ref{eq:E_DGM}), 
the energy of two  free Giant Magnons in the non-relativistic 
regime ($\delta,|Q^{(1)}|,|Q^{(2)}|<<1$) is given as, 
\begin{equation}\label{eq:energy}
E_{1+2}=2M+\frac{1}{2}M[\omega^{(1)}]^2+
\frac{1}{2}M[\omega^{(2)}]^2-
\frac{1}{2}M[V^{(1)}]^2+\frac{1}{2}M[V^{(2)}]^2,
\end{equation}
where the effective mass (and moment of inertia) is 
$M=M(0)=\sqrt{\lambda}/\pi$.
The leading large-distance interaction between the two solitons 
can be incorporated by  adding the two-body potential $\mathcal{V}$ 
introduced above. Finally, the effective Hamiltonian 
describing the interaction of two dyonic giant
magnons, in the regime where they are well separated and slow moving
can be written as, 
\begin{equation}\label{eq:hamiltonian}
\mathcal{H}=2M-\frac{[P^{(1)}]^2}{2m}-\frac{[P^{(2)}]^2}{2m}+
\frac{[Q^{(1)}]^2}{2m}+\frac{[Q^{(2)}]^2}{2m}+
\mathcal{V}(|\Delta
X|,\Theta^{(1)},\Theta^{(2)}),
\end{equation}
where $P^{(a)}$ and $Q^{(a)}$ the canonical conjugate momenta to $X^{(a)}$
and $\Theta^{(a)}$ respectively. The corresponding Poisson brackets are, 
\begin{eqnarray}
\{Q^{(a)},Q^{(b)}\}=\{P^{(a)},P^{(b)}\}=\{Q^{(a)},P^{(b)}\}=0,\\
\{X^{(a)},X^{(b)}\}=\{\Theta^{(a)},\Theta^{(b)}\}=\{X^{(a)},\Theta^{(b)}\}=0,\\
\{P^{(a)},X^{(b)}\}=\delta_{ab},\quad
\{Q^{(a)},\Theta^{(b)}\}=\delta_{ab},\quad a,b=1,2,
\end{eqnarray}
By construction, equations of motion following 
from the Hamiltonian (\ref{eq:hamiltonian}) coincide with 
(\ref{eq:Force}-\ref{eq:Torque}), and the expression 
provides the expected energy (\ref{eq:energy}) when
$d\rightarrow\infty$. 

In \cite{DHM} a toy model for the dynamics of Giant Magnons was
presented. An attractive feature of the model was that the resulting
S-matrix for magnon scattering had a similar analytic structure to the
exact S-matrix known from the asymptotic Bethe ansatz. 
For two magnons aligned in the same $S^{3}$ subspace of $S^{5}$, the 
toy model Hamiltonian can be cast in the form
\begin{equation}
\mathcal{H}_{DHM}=2M-\Re\biggl\{\frac{2\pi_1^2}{m}+\frac{2\pi_2^2}{m}+
\frac{A}{\sinh^2(z_1-z_2)}
\biggr\},\end{equation}
for complex phase space variables $(\pi_a,z_a)\in\mathbb C^2$,
with $a=1,2$. By taking $z_a=\frac12(X^{(a)}+i\Theta^{(a)})$,
$\pi_a=\frac12(P^{(a)}+iQ^{(a)})$ and going to the limit of large
separation $d=|X^{(1)}-X^{(2)}|>>1$, the Hamiltonian reduces to
\begin{equation}
\mathcal{H}_{DHM}\approx 2M
-\frac{[P^{(1)}]^2}{2M}-\frac{[P^{(2)}]^2}{2M}+\frac{[Q^{(1)}]^2}{2M}+
\frac{[Q^{(2)}]^2}{2M}+4A\,\cos(\Delta\Theta)\,e^{-\Delta
X}.
\end{equation}
In the case $A=-M=-\sqrt{\lambda}/\pi$, we see that $H_{DGM}$
coincides with the effective Hamiltonian $\mathcal{H}$ as given in 
(\ref{eq:hamiltonian}) when the solitons have parallel angular
momenta, i.e., $\vec J_1^\pm\cdot \vec J_2^\pm = J_1J_2=Q_1Q_2$.


\section*{Appendix}

\subsection*{A basis for $\mathfrak{su}(2)\oplus\mathfrak{su}{2}$}

Through this paper we have used the following basis $\{\t_a^\pm\}$
for the Lie algebra
$\mathfrak{su}(2)\oplus\mathfrak{su}(2)$,
\[
\t_1^-=\begin{pmatrix}
0 & 1 & 0 & 0\\
-1 & 0 & 0 & 0\\
0 & 0 & 0 & 1\\
0 & 0 & -1 & 0
\end{pmatrix},\quad
\t_2^-=\begin{pmatrix}
0 & 0 & 0 & 1\\
0 & 0 & 1 & 0\\
0 & -1 & 0 & 0\\
-1 & 0 & 0 & 0
\end{pmatrix},\quad
\t_3^-=\begin{pmatrix}
0 & 0 & 1 & 0\\
0 & 0 & 0 & -1\\
-1 & 0 & 0 & 0\\
0 & 1 & 0 & 0
\end{pmatrix},
\]
\[
\t_1^+=\begin{pmatrix}
0 & -1 & 0 & 0\\
1 & 0 & 0 & 0\\
0 & 0 & 0 & 1\\
0 & 0 & -1 & 0
\end{pmatrix},\quad
\t_2^+=\begin{pmatrix}
0 & 0 & 0 & 1\\
0 & 0 & -1 & 0\\
0 & 1 & 0 & 0\\
-1 & 0 & 0 & 0
\end{pmatrix},\quad
\t_3^+=\begin{pmatrix}
0 & 0 & -1 & 0\\
0 & 0 & 0 & -1\\
1 & 0 & 0 & 0\\
0 & 1 & 0 & 0
\end{pmatrix}.
\]
These obey the relations
\[[\t^\pm_i,\t^\pm_j]=\mp\epsilon_{ijk} \t^\pm_k,\quad
\mathrm{tr}(\t^+_i\t^-_j)=0,\quad\mathrm{tr}(\t^\pm_i,\t^\pm_j)=-4\delta_{ij},\quad(i,j,k=1,2,3)\]

\subsection*{Finding the 2-DGM Solution}

The string action (\ref{eq:action}), when restricted to the sector
$\mathbb{R}\times S^5$, with $Y_0=t$, can be rewritten as the action
for a sigma model on the coset space
$SU(4)/Sp(2)\approx S^5$.  An element of this coset will
obey
\[\g^\dagger\g=\id,\]
\[\J\g\J^{-1}=\g^T,\]
where
\[\J=\begin{pmatrix} 0 & \id \\ -\id & 0 \end{pmatrix}.\]

A convenient parametrisation that makes
contact with (\ref{eq:action},\ref{eq:Zs}) is
\[\g=\begin{pmatrix}
Z_1 & Z_2 & 0 & Z_3 \\
-\bar Z_2 & \bar Z_1 & -Z_3 & 0\\
0 & \bar Z_3 & Z_1 & -\bar Z_2  \\
-\bar Z_3 & 0 & Z_2 & \bar Z_1
\end{pmatrix}.\]

The first coset condition becomes
\[|Z_1|^2+|Z_2|^2+|Z_3|^2=1,\]
and the second is trivially verified.

A simple solution for the equations of motion is
\[\g_0=\diag(e^{it},e^{-it},e^{it},e^{-it}).\]
This corresponds to a point particle moving along the equator of
$S^5$: the BMN string (\ref{eq:BMN}).

The correspondent wave-function can then be computed,
\[\Psi_0=\diag(e^{iZ},e^{-iZ},e^{iZ},e^{-iZ}),\]
with
\[Z(\lambda)=\frac{x_+}{\lambda+1}+\frac{x_-}{\lambda-1},\quad x_{\pm}=\frac12
(x\pm
t),\]
$(x,t)$ the worldsheet string coordinates, and $\lambda$ a complex
parameter.

In the description of the solitons it is convenient to define the
coordinates $(u,v)$ by
\[u=i[Z(\lambda)-Z(\bar \lambda)],\quad v=Z(\lambda)+Z(\bar \lambda)-t.\]
If we write $\lambda=r e^{ip/2}$, and make $p=\pi-\delta$ these become
\begin{equation}\label{uvs}
 u=\frac{2 r (1 + r^2)\cos\bigl(\frac{\delta}2\bigr)}{1 + r^4 + 2 r^2
\cos(\delta)}[x-V\,t],\quad v=\frac{2 r
(r^2-1)\sin\bigl(\frac{\delta}2\bigr)}{1 + r^4 + 2 r^2
\cos(\delta)}[x-\frac{t}{V}],
\end{equation}
with
\[V=\frac{2r}{r^2+1}\sin\biggl(\frac{\delta}2\biggr).\]
These are the functions (\ref{eq:u}-\ref{eq:v}) defined on the main
text.

Using the dressing method, the 1-soliton wave function can be computed
by dressing the vacuum wave function,
\[\Psi_1=\chi_1\Psi_0,\]
with the dressing factor
\[\chi_1=\id+\frac{\lambda_1-\bar
\lambda_1}{\lambda-\lambda_1}\P_0+\frac{1/\bar \lambda_1-1/
\lambda_1}{\lambda-1/\bar \lambda_1}\Q_0,\]
where $\lambda_1=r_1 e^{i\frac{p_1}2}$ is the complex parameter that
classifies the properties of the soliton solution (such as its
momentum and angular momentum). $\P_0$ and
$\Q_0$ are projectors that are determined from the vacuum wave-function:
\[\P_1=\frac{\ap\otimes\bp}{\ap\cdot\bp},\quad
\Q_1=\frac{\aq\otimes\bq}{\aq\cdot\bq}.\]
\[\ap=\Psi_0(\bar \lambda_1)w_1,\quad\bp=w_1^\dagger [\Psi_0(\bar
\lambda_1)]^\dagger,\]
\[\aq=\Psi_0(\lambda_1^{-1})\J\bar
w_1,\quad\bq=\J^{-1}[\Psi(\lambda_1^{-1})]^\dagger.\]
$w_1$ is a $4$-vector that determines the polarisation of the giant
magnon. The soliton solutions (\ref{eq:DGMs}) with parameters
(\ref{eq:DGMs_par}) can be generated separately from the polarisations
\begin{equation}\label{eq:polarizations}
w_1 =\begin{pmatrix} i & 1 & 0
&0\end{pmatrix}^T,\quad\quad
w_2=\begin{pmatrix}
ie^{-D/2+i(\psi_++\psi_--\theta_0)/2}\cos\alpha_-\\
e^{D/2+i(-\psi_++\psi_-+\theta_0)/2}\cos\alpha_+\\
-ie^{-D/2+i(-\psi_+-\psi_--\theta_0)/2}\sin\alpha_-\\
e^{D/2+i(\psi_+-\psi_-+\theta_0)/2}\sin\alpha_+
\end{pmatrix},
\end{equation}
by taking $\g_1=\Psi_1[\lambda=0]$.

We wish to identify a 2-DGM solution that at $t=0$ resembles, near the
vicinity of each soliton, the direct sum of the solutions
(\ref{eq:DGMs}). If we naively pick the solution
\[\g_2=\chi_2[\lambda=0;w_2]\,\chi_1[\lambda=0;w_1]\,\g_1,\]
we find out that separation between the solitons is no longer
$d$ but that it receives a leading order correction due to a
time delay in the scattering. In such solution, the solitons are not
positioned at $t=0$ as those in (\ref{eq:DGMs}-\ref{eq:DGMs_par}).
Further more, the solution $\g_2$ appears to be rotated, the first soliton
not being fixed at $Z_3=0$. By changing the
polarisations and inducing a global rotation in $\g_2$, we can
generate a two soliton solution $\tilde\g_2$ that resembles the direct sum of
the solitons
(\ref{eq:DGMs}-\ref{eq:DGMs_par}) at $t=0$, in the neighbourhood of the
centre of each DGM. This solution can be formally constructed from the
polarizations:
\begin{equation}\label{eq:polarizations_corrected}
\tilde w_1 =\begin{pmatrix} ie^{-\frac{\Delta_-}2} &
e^{\frac{\Delta_-}2} & 0
&0\end{pmatrix}^T,\quad\quad
\tilde w_2=\begin{pmatrix}
ie^{-(D-\Delta_+)/2+i(\psi_++\psi_--\theta_0)/2}\cos\alpha_-\\
e^{(D-\Delta_+)/2+i(-\psi_++\psi_-+\theta_0)/2}\cos\alpha_+\\
-ie^{-(D-\Delta_+)/2+i(-\psi_+-\psi_--\theta_0)/2}\sin\alpha_-\\
e^{(D-\Delta_+)/2+i(\psi_+-\psi_-+\theta_0)/2}\sin\alpha_+
\end{pmatrix},
\end{equation}
with $\Delta_\pm$ given by
\begin{equation}
\Delta_\pm=\pm\frac12\log\biggl(\frac{A}{B_\pm}\biggr),
\end{equation}
and
\begin{equation}
A=(r_1+r_2)^2 (1+r_1 r_2)^2,
\end{equation}
\begin{multline}
B_\pm=r_2^2+r_1^4 r_2^2+r_1
(r_2+r_2^3)+r_1^3 (r_2+r_2^3)+r_1^2 (1+r_2^4)
\\
-r_1 r_2 \biggl[2 (-1+r_1^2) (-1+r_2^2)
\cos(2 \alpha_\pm)\cos^2\bigl(\frac{\delta}2\bigr)+(1+4 r_1 r_2+r_2^2+r_1^2
(1+r_2^2))
\cos\delta\biggr].
\end{multline}
These expressions simplify to (\ref{eq:Deltas}) in the
non-relativistic regime.
The solution will then be given, up to a global rotation, by
\begin{equation}
\tilde \g_2 = \chi_{\tilde w_2}(0)\chi_{\tilde w_1}(0)\g_0.
\end{equation}
We don't need to determine the explicit form of $\tilde\g_2$, only its
NLO corrections (the LO correction is compensated already by our
particular choice of the polarizations).

Let us turn back to $\g_2$ again. For well separated solitons, $D>>1$, we can
expand
the dressing factor in powers of $e^{-D}$
\begin{equation}
\label{eq:chi_expan}
\chi_{2}=\chi^{(0)}_{2}+ \chi^{(1)}_{2}e^{-D}+\mathcal{O}[e^{-2D}].
\end{equation}
We note that the vectors $\vec a_2, \vec b_2$ exhibit the following
dependence on $D$,
\[\vec a_2 = \vec a_2^+ e^{\frac{D}2}+\vec a_2^- e^{-\frac{D}2},\quad
\vec b_2 = \vec b_2^+ e^{\frac{D}2}+\vec b_2^- e^{-\frac{D}2}.\]
By setting
\[\P_2=\frac{\N_{\P 2}}{\D_{\P 2}},\quad\Q_2=\frac{\N_{\Q 2}}{\D_{\Q 2}},\]
this translates into
\[\N_2\equiv\vec a_2\otimes\vec b_2=\N_2^{+}e^{D}+\N_2^0+\N_2^- e^{-D},\]
\[\D_2\equiv\vec a_2\cdot\vec b_2=\D_2^{+}e^{D}+\D_2^0+\D_2^-
e^{-D},\]
where
\[\N_2^\pm \equiv \vec a_2^\pm\otimes \vec b_2^\pm,\quad
\N_2^0=\vec a_2^+\otimes \vec b_2^-+\vec a_2^-\otimes \vec b_2^+,\]
\[\D_2^\pm \equiv \vec a_2^\pm\cdot \vec b_2^\pm,\quad
\N_2^0=\vec a_2^+\cdot \vec b_2^-+\vec a_2^-\cdot \vec b_2^+.\]
By expanding in $e^{-D}$ the projectors, we have that
\[\P_{2}^{(0)}=\frac{\N_{\P 2}^+}{\D_{\P 2}^+},\quad
\Q_{2}^{(0)}=\frac{\N_{\Q 2}^+}{\D_{\Q 2}^+},\quad\text{at LO}\]
\[\P_{2}^{(1)}=\frac1{\D_{\P 2}^+}\Bigl(\N^0_{\P
2}-\frac{\D^0_{\P 2}}{\D^+_{\P 2}}\N^+_{\P 2}\Bigr),\quad
\Q_{2}^{(1)}=\frac1{\D_{\Q 2}^+}\Bigl(\N^0_{\Q
2}-\frac{\D^0_{\Q 2}}{\D^+_{\Q 2}}\N^+_{\Q
2}\Bigr),\quad\text{at NLO}\]
From where the terms in (\ref{eq:chi_expan}) can be computed. The LO
correction will be responsible for a shift in the positions of the
solitons at time $t=0$, in relation to the direct sum. The shifts will
be given by
\begin{equation}
\Delta X_1 = \tan\bigl(\frac\delta{2}\bigr)\frac{\Delta_-}{\gamma_1^2 V_1}
\Delta X_2 = -\tan\bigl(\frac\delta{2}\bigr)\frac{\Delta_+}{\gamma_2^2 V_2}
\end{equation}
where $\gamma_j=(1-V_j^2)^{\frac{1}2}$, and $\Delta_\pm$ can be
approximated, in the non-relativist regime $\delta<<1$, by
\begin{equation}\label{eq:Deltas}
\exp(\pm 2\Delta_\pm)\approx \frac{(m_1+m_2)^2}{(\vec J_1^\pm-\vec
J_2^\pm)^2},
\end{equation}
$\vec J_j^\pm$ being the angular momenta of the free solutions
(\ref{eq:DGMs}). The time delay is proportional to
$\Delta_+-\Delta_-$. As previously mentioned, the polarisations
(\ref{eq:polarizations}) should be modified to accommodate these
changes. This is easily done by the change $D\rightarrow D-\Delta_+$
on $w_2$ and by taking $w_1=(i
e^{-\frac{\Delta_-}2},e^{\frac{\Delta_-}2},0,0)$. This justifies the
selection of the polarizations $\tilde w_1$ and $\tilde w_2$ in the
construction of $\tilde\g_2$.

Not only the positions of the solitons have been changed due to LO
corrections, but the solution appears rotated as well. The relative
orientation of the two solitons is kept, but a global rotation was
induced by the application of the dressing method. To undo this, we
need to choose the rotation that fixes $Z_3=0$ at LO in the
neighbourhood of the origin. The rotation is easily found to be
\[\tilde Z_{2}^{(0)}=\bar C_{2} Z_{2}^{(0)}+\bar C_{3} Z_{3}^{(0)},\]
\[\tilde Z_{3}^{(0)}=C_{3} Z_{2}^{(0)}-C_{2} Z_{3}^{(0)},\]
with
\begin{align*}
C_{2} &=e^{-\Delta_-}\frac{[r_2 - r_1^2 r_2 + r_1 (-1 + r_2^2) \cos(2
\alpha_+)] (1 +
    \cos\delta_1) + i (r_1 + r_2) (1 + r_1 r_2)\sin\delta_1}
{(1 + e^{i \delta_1}) (r_1 + r_2) (1 + r_1 r_2)},\\
C_{3} &=e^{-\Delta_-}\frac{e^{-i (\delta_1 + \psi_+)}
(1 + e^{
   i \delta_1}) r_1 (-1 + r_2^2)\cos\alpha_+\sin\alpha_+}{(r_1 + r_2) (1 + r_1
r_2)},
\end{align*}
with $|C_2|^2+|C_3|^2=1$. $Z_1$ remains unmodified. With this rotation
in place (applied to the NLO as well), and with the polarisation shifts,
the 2-DGM solution thus obtained can be approximated at LO, in the
neighbourhood of each soliton,
by the sum of the two free solutions (\ref{eq:DGMs}). In particular,
for $\delta Z_2^{(1)}$ we find\footnote{we have absorbed $\theta_0$ in
$v_2$ in these expressions.}, 
\begin{multline*}
\delta Z_2^{(1)}=-\frac{ie^{-u_2-i (v_2+\alpha_-+\alpha_++\delta+i
\psi_+)} (1 +e^{i \delta})
}{4 (1
+e^{2 u_1})^2 (r_1+r_2)^2(1+r_1r_2)^2}
\\
\times \biggl[
-e^{2 i \psi_+}
(r_1+r_2)^2
+e^{2 i (\alpha_-+\psi_+)} (r_1+r_2)^2
+e^{2 i (\alpha_++\psi_+)} (r_1+r_2)^2
\\
-e^{2 i
(\alpha_-+\alpha_++\psi_+)} (r_1+r_2)^2
-e^{2 (u_1+i
(v_1+v_2))} r_1 r_2 (r_1+r_2)^2
\\
+e^{2 (u_1+i
(v_1+v_2+\alpha_-))} r_1 r_2 (r_1+r_2)^2
+e^{2 (u_1+i
(v_1+v_2+\alpha_+))} r_1 r_2 (r_1+r_2)^2
\\
-e^{2 (u_1+i
(v_1+v_2+\alpha_-+\alpha_+))} r_1 r_2 (r_1+r_2)^2
-e^{2
(u_1+i (v_1+v_2+\delta))} r_1 r_2 (r_1+r_2)^2
\\
-2 e^{2
u_1+i (2 v_1+2 v_2+\delta)} r_1 r_2 (r_1+r_2)^2
+e^{2
(u_1+i (v_1+v_2+\alpha_-+\delta))} r_1 r_2 (r_1+r_2)^2
\\
+2
e^{2 u_1+i (2 v_1+2 v_2+2 \alpha_-+\delta)} r_1 r_2
(r_1+r_2)^2
+e^{2 (u_1+i (v_1+v_2+\alpha_++\delta))} r_1
r_2 (r_1+r_2)^2
\\
-e^{2 (u_1+i
(v_1+v_2+\alpha_-+\alpha_++\delta))} r_1 r_2 (r_1+r_2)^2
+2 e^{2
u_1+i (2 v_1+2 v_2+2 \alpha_++\delta)} r_1 r_2
(r_1+r_2)^2
\\
-2 e^{2 u_1+i (2 v_1+2 v_2+2 \alpha_-+2
\alpha_++\delta)} r_1 r_2 (r_1+r_2)^2+2 e^{i (\delta+2 \psi_+)}
r_1 r_2 (r_1+r_2)^2
\\
-2 e^{i (2 \alpha_-+\delta+2 \psi_+)} r_1 r_2
(r_1+r_2)^2
-2 e^{i (2 \alpha_++\delta+2 \psi_+)} r_1 r_2
(r_1+r_2)^2
+2 e^{i (2 \alpha_-+2 \alpha_++\delta+2 \psi_+)} r_1
r_2 (r_1+r_2)^2
\\
-e^{2 i (\delta+\psi_+)} r_1^2 r_2^2 (r_1+r_2)^2
+e^{2 i
(\alpha_-+\delta+\psi_+)} r_1^2 r_2^2 (r_1+r_2)^2+
\cdots
\end{multline*}
\begin{multline}\label{eq:deltaZ2}
\cdots
-e^{2
u_1+2 i \psi_+} (r_1+r_2)^2 (1+r_1 r_2)
+e^{2
(u_1+i (\alpha_-+\psi_+))} (r_1+r_2)^2 (1+r_1 r_2)
+e^{2
(u_1+i (\alpha_++\psi_+))} (r_1+r_2)^2 (1+r_1 r_2)
\\
-e^{2
(u_1+i (\alpha_-+\alpha_++\psi_+))} (r_1+r_2)^2
(1+r_1 r_2)
\\
-e^{2 (u_1+i (\delta+\psi_+))} r_1 r_2
(r_1+r_2)^2 (1+r_1 r_2)
+e^{2 (u_1+i
(\alpha_-+\delta+\psi_+))} r_1 r_2 (r_1+r_2)^2 (1+r_1 r_2)
\\
+e^{2
(u_1+i (\alpha_++\delta+\psi_+))} r_1 r_2 (r_1+r_2)^2
(1+r_1 r_2)-e^{2 (u_1+i
(\alpha_-+\alpha_++\delta+\psi_+))} r_1 r_2 (r_1+r_2)^2 (1+r_1
r_2)
\\
+e^{2 i (v_2+\delta+\psi_+)} r_1^2 (1+r_1 r_2)^2
+e^{2 i
(v_2+\alpha_-+\delta+\psi_+)} r_1^2 (1+r_1 r_2)^2
\\
+e^{2 i
(v_2+\alpha_++\delta+\psi_+)} r_1^2 (1+r_1 r_2)^2
+e^{2 i
(v_2+\alpha_-+\alpha_++\delta+\psi_+)} r_1^2 (1+r_1 r_2)^2
\\
+e^{2
u_1+2 i v_1} r_1 r_2 (1+r_1 r_2)^2
+e^{2 (u_1+i
(v_1+\alpha_-))} r_1 r_2 (1+r_1 r_2)^2
\\
+e^{2 (u_1+i
(v_1+\alpha_+))} r_1 r_2 (1+r_1 r_2)^2
+e^{2 (u_1+i
(v_1+\alpha_-+\alpha_+))} r_1 r_2 (1+r_1 r_2)^2
\\
+e^{2
(u_1+i (v_1+\delta))} r_1 r_2 (1+r_1 r_2)^2+2 e^{2
u_1+i (2 v_1+\delta)} r_1 r_2 (1+r_1 r_2)^2
+e^{2
(u_1+i (v_1+\alpha_-+\delta))} r_1 r_2 (1+r_1 r_2)^2
\\
+2
e^{2 u_1+i (2 v_1+2 \alpha_-+\delta)} r_1 r_2 (1+r_1
r_2)^2
+e^{2 (u_1+i (v_1+\alpha_++\delta))} r_1 r_2
(1+r_1 r_2)^2
\\
+e^{2 (u_1+i
(v_1+\alpha_-+\alpha_++\delta))} r_1 r_2 (1+r_1 r_2)^2+2 e^{2
u_1+i (2 v_1+2 \alpha_++\delta)} r_1 r_2 (1+r_1 r_2)^2
\\
+2
e^{2 u_1+i (2 v_1+2 \alpha_-+2 \alpha_++\delta)}
r_1 r_2 (1+r_1 r_2)^2-
\\
-2 e^{i (2 v_2+\delta+2 \psi_+)} r_1 r_2 (1+r_1
r_2)^2-2 e^{i (2 v_2+2 \alpha_-+\delta+2 \psi_+)} r_1 r_2 (1+r_1
r_2)^2
\\
-2 e^{i (2 v_2+2 \alpha_++\delta+2 \psi_+)} r_1 r_2 (1+r_1
r_2)^2-2 e^{i (2 v_2+2 \alpha_-+2 \alpha_++\delta+2 \psi_+)} r_1
r_2 (1+r_1 r_2)^2
\\
+e^{2 i (v_2+\psi_+)} r_2^2 (1+r_1 r_2)^2
+e^{2 i
(v_2+\alpha_-+\psi_+)} r_2^2 (1+r_1 r_2)^2
+e^{2 i
(v_2+\alpha_++\psi_+)} r_2^2 (1+r_1 r_2)^2
\\
+e^{2 i
(v_2+\alpha_-+\alpha_++\psi_+)} r_2^2 (1+r_1 r_2)^2
+e^{2
(u_1+i (v_2+\delta+\psi_+))} r_1 (r_1+r_2) (1+r_1
r_2)^2
\\
+e^{2 (u_1+i (v_2+\alpha_-+\delta+\psi_+))}
r_1 (r_1+r_2) (1+r_1 r_2)^2
+e^{2 (u_1+i
(v_2+\alpha_++\delta+\psi_+))} r_1 (r_1+r_2) (1+r_1 r_2)^2
\\
+e^{2
(u_1+i (v_2+\alpha_-+\alpha_++\delta+\psi_+))}
r_1 (r_1+r_2) (1+r_1 r_2)^2
+e^{2 (u_1+i (v_2+\psi_+))} r_2
(r_1+r_2) (1+r_1 r_2)^2
\\
+e^{2 (u_1+i
(v_2+\alpha_-+\psi_+))} r_2 (r_1+r_2) (1+r_1 r_2)^2
+e^{2
(u_1+i (v_2+\alpha_++\psi_+))} r_2 (r_1+r_2) (1+r_1
r_2)^2
\\
+e^{2 (u_1+i (v_2+\alpha_-+\alpha_++\psi_+))}
r_2 (r_1+r_2) (1+r_1 r_2)^2
+e^{2 u_1+i (\delta+2
\psi_+)} (r_1+r_2)^2(1+r_1r_2)^2
\\
+e^{4 u_1+i
(\delta+2 \psi_+)} (r_1+r_2)^2(1+r_1r_2)^2-e^{2
u_1+i (2 v_2+\delta+2 \psi_+)} (r_1+r_2)^2(1+r_1r_2)^2
\\
-e^{4 u_1+i (2 v_2+\delta+2 \psi_+)}
(r_1+r_2)^2(1+r_1r_2)^2
\\
-e^{2 u_1+i (2
\alpha_-+\delta+2 \psi_+)} (r_1+r_2)^2(1+r_1r_2)^2-e^{4
u_1+i (2 \alpha_-+\delta+2 \psi_+)} (r_1+r_2+r_1^2
r_2+r_1 r_2^2)^2
\\
-e^{2 u_1+i (2 v_2+2 \alpha_-+\delta+2
\psi_+)} (r_1+r_2)^2(1+r_1r_2)^2-e^{4 u_1+i (2 v_2+2
\alpha_-+\delta+2 \psi_+)} (r_1+r_2)^2(1+r_1r_2)^2
\\
-e^{2
u_1+i (2 \alpha_++\delta+2 \psi_+)} (r_1+r_2+r_1^2
r_2+r_1 r_2^2)^2
\\
-e^{4 u_1+i (2 \alpha_++\delta+2
\psi_+)} (r_1+r_2)^2(1+r_1r_2)^2-e^{2 u_1+i (2 v_2+2
\alpha_++\delta+2 \psi_+)} (r_1+r_2)^2(1+r_1r_2)^2
\\
-e^{4
u_1+i (2 v_2+2 \alpha_++\delta+2 \psi_+)}
(r_1+r_2)^2(1+r_1r_2)^2
\\
+e^{2 u_1+i (2 \alpha_-+2
\alpha_++\delta+2 \psi_+)} (r_1+r_2)^2(1+r_1r_2)^2
+e^{4
u_1+i (2 \alpha_-+2 \alpha_++\delta+2 \psi_+)}
(r_1+r_2)^2(1+r_1r_2)^2
\\
-e^{2 u_1+i (2 v_2+2
\alpha_-+2 \alpha_++\delta+2 \psi_+)} (r_1+r_2)^2(1+r_1r_2)^2
\\
-e^{4 u_1+i (2 v_2+2 \alpha_-+2
\alpha_++\delta+2 \psi_+)} (r_1+r_2)^2(1+r_1r_2)^2\biggr].
\end{multline}

It is straightforward to check that, for $\delta<<1$ and $u_1\approx
0$,
\begin{multline}\label{eq:frac}
\biggl(\frac{\delta Z_2^{(1)}}{Z_2^{(1)}}\biggr)_{x=\omega^{(1)}t}=
 4i\,r_1 r_2\biggl\{\frac1{(r_1 + r_2)^2}\cos \alpha_-\cos\alpha_+\sin(v_1 -
v_2 -\psi_+) \\+ \frac1{(1 + r_1 r_2)^2} \sin\alpha_- \sin\alpha_+
\sin(v_1 + v_2 - \psi_+)\biggr\}
\end{multline}


\end{document}